\documentclass[preprint,aps,prd,12pt,preprintnumbers,nofootinbib,superscriptaddress]{revtex4-2}
\usepackage{xcolor}
\usepackage{braket}
\usepackage{amssymb,amsmath,amsfonts,comment,latexsym,graphicx,epstopdf,float}
\usepackage{color}
\usepackage{fancyvrb}
\usepackage{chngcntr}
\usepackage{etoolbox}
\usepackage{ulem}
\usepackage{array}
\usepackage{bbold}
\usepackage{subfigure}
\usepackage{slashed}
\usepackage{multirow}

\usepackage[colorlinks=true,citecolor=blue,linkcolor=red,filecolor=purple,urlcolor=magenta]{hyperref}

\usepackage{float}

\newcommand{\be}{\begin{equation}}
\newcommand{\ee}{\end{equation}}
\newcommand{\ba}{\begin{eqnarray}}
\newcommand{\ea}{\end{eqnarray}}

\newcommand{\grts}{\raise.3ex\hbox{$>$\kern-.75em\lower1ex\hbox{$\sim$}}}
\newcommand{\lets}{\raise.3ex\hbox{$<$\kern-.75em\lower1ex\hbox{$\sim$}}}

\usepackage{color}
\usepackage{amssymb,amsmath,amsfonts}
\usepackage{epstopdf} 
\usepackage{braket}

\usepackage{autobreak}

\begin{document}
%
%
\title{\vspace*{0.5in} 
Implications of asymptotic safety in two minimal Z-prime models
\vskip 0.1in}
\author{Christopher D. Carone}\email[]{cdcaro@wm.edu}
\affiliation{High Energy Theory Group, Department of Physics, William \& Mary, Williamsburg, VA 23187-8795, USA}
\date{December 12, 2025}
%
%
\begin{abstract}
We consider the implications of asymptotic safety on two U(1) gauge extensions of the standard model that are minimal in the sense that anomaly cancellation only requires the presence of right-handed neutrinos.   We study the UV fixed points of the gauge couplings taking into account kinetic mixing between hypercharge and the new U(1) gauge field.   We consider the possibility that the top-bottom mass splitting originates from the effect of differing gauge charges on the nontrivial fixed point values of their respective Yukawa couplings and assess the impact of the extended gauge symmetry on the viability of this picture.   
\end{abstract}

\maketitle
\newpage

\section{Introduction} \label{sec:intro}
Extensions of the standard model often involve new couplings that are not strongly constrained by the available experimental data.  This is the case, for example, if the scale of new physics is beyond the reach of current collider experiments and the new physics does not violate any fundamental symmetries.   For this reason, theories with minimal additional particle content are of particular interest, especially if the couplings are constrained by some simple organizing principle.   New symmetries or specific ultraviolet (UV) boundary conditions can provide such restrictions.  For example, the requirement that gauge couplings unify at a particular high scale~\cite{Georgi:1974yf} is a UV boundary condition that reduces the space of free parameters, leading to a more predictive low-energy theory.

Asymptotic safety~\cite{weinberg0} provides a different UV boundary condition that can reduce the space of free parameters in a beyond-the-standard-model (BSM) theory~\cite{Bond:2017wut,Wang:2015sxe,Grabowski:2018fjj,Hiller:2019mou,Kowalska:2020zve,Kowalska:2020gie,Bause:2021prv,Chikkaballi:2022urc,Boos:2022jvc,Boos:2022pyq,Chikkaballi:2025pnw}.   In the asymptotical safety paradigm, one assumes that all the dimensionless couplings in the theory can be run to infinitely high energy without any reaching a Landau pole.  (This includes dimensionless couplings that can be formed by multiplying a coupling by a power of the renormalization scale.) The subset of the space of couplings where this statement holds true is called the ultraviolet critical surface.   A theory is not considered to be physically sensible if the couplings do not live on this surface.   As a result, a theory that is nonrenormalizable, with an infinite number of Lagrangian parameters, may nonetheless remain predictive since the UV critical surface may be finite dimensional.   This statement is the basis of Weinberg's original observation that asymptotic safety may render a quantum field theory of gravity predictive, even if it is nonrenormalizable by power counting arguments~\cite{weinberg0}.   Using the functional renormalization group approach~\cite{Wetterich:1992yh,Morris:1993qb}, subsequent nonperturbative studies of gravity~\cite{Reuter:1996cp,Lauscher:2001ya,Reuter:2001ag,Litim:2003vp,Codello:2006in,Machado:2007ea,Codello:2008vh,Benedetti:2009rx,Dietz:2012ic,Falls:2014tra}  have supported the possibility that gravity may be asymptotically safe.  For a review of asymptotic safety, see Refs.~\cite{Percacci:2007sz,Eichhorn:2018yfc}.

Asymptotic safety must also extend to the matter sector of the theory~\cite{Percacci:2002ie,Daum:2009dn,Zanusso:2009bs,Folkerts:2011jz,Dona:2013qba,Meibohm:2015twa,Oda:2015sma,Eichhorn:2016esv,Christiansen:2017cxa,Eichhorn:2017eht}.  Without including gravity, the standard model is not asymptotically safe due to the running of the hypercharge gauge coupling, which reaches a Landau pole above the Planck scale.   However, functional renormalization group studies of Abelian gauge theories suggest that gravitational physics can alter that conclusion~\cite{Harst:2011zx,Christiansen:2017gtg,Eichhorn:2017lry}.   For a gauge coupling $g$, the leading effect of gravity is approximated by a term that is linear in the coupling~\cite{Folkerts:2011jz,Eichhorn:2016esv,Christiansen:2017gtg,Eichhorn:2017eht}
\begin{equation}
\beta(g) = \frac{1}{(4 \pi)^2} \, \beta^{(1)}(g) - \theta(\mu-M_{\rm Pl}) \, f_g \, g \,\,\, ,
\label{eq:linterm}
\end{equation}
where $M_{\rm Pl}$ is the Planck scale, $\theta$ is the Heaviside step function, and $f_g$ parameterizes the gravitational physics.  In this paper, we will work consistently with one-loop renormalization group equations in the $\overline{\rm MS}$ scheme.  Note that $f_g$ is universal for all the gauge couplings; similar linear correction terms that turn on at the Planck scale are present for the Yukawa couplings ($f_y$) and the couplings in the scalar potential ($f_\lambda$).  Calculations of $f_g$ involve significant uncertainties, from the truncation of the matter-gravity action and from renormalization scheme dependence, but typically lead to $f_g$ being nonnegative~\cite{Folkerts:2011jz}.  This is a standard assumption in the phenomenological literature, which we adopt here, and corresponds to schemes that break certain classical symmetries that would otherwise give $f_g=0$~\cite{Folkerts:2011jz}.   We do not restrict the sign of $f_y$; our phenomenologically viable examples in Sec.~\ref{sec:conseq} include solutions with either sign.  Given the form of Eq.~(\ref{eq:linterm}), $f_g$ may be chosen so that $\beta(g)$ vanishes at $M_{\rm Pl}$.  In this case, $g$ reaches the nontrivial UV fixed point $g_* = g(M_{\rm Pl})$; for larger choices of $f_g$, the coupling flows to a Gaussian fixed point, {\it i.e.}, where $g_* =0$.    For a UV complete theory, we will use the term asymptotically safe (rather than asymptotically free) if at least one coupling reaches a nontrivial UV fixed point.   Scenarios in which multiple couplings are required to reach nontrivial fixed points lead to interesting restrictions on the low-energy theory.   This fact has been used previously in the context of BSM model building~\cite{Bond:2017wut,Wang:2015sxe,Grabowski:2018fjj,Hiller:2019mou,Kowalska:2020zve,Kowalska:2020gie,Bause:2021prv,Chikkaballi:2022urc,Boos:2022jvc,Boos:2022pyq,Chikkaballi:2025pnw}.
 
Our present interest is the implications of asymptotic safety on minimal U(1) gauge extensions of the standard model (referred also as $Z$-prime models).  In the author's previous work~\cite{Boos:2022jvc}, Boos {\it et al.}, studied this issue in the context a model with gauged baryon number, demonstrating how the kinetic mixing parameter (which is normally an undetermined coupling) could be meaningfully constrained by the requirement that the Abelian gauge couplings reach nontrivial UV fixed points.  The same authors went on to consider how dark matter could be incorporated into the model~\cite{Boos:2022pyq}.   While the possibility of gauged baryon number has been well studied in other contexts~\cite{gbn}, the theory is not the most elegant, requiring a substantial sector of additional fermion to cancel anomalies ({\it i.e.,} once that are vector-like under the standard model gauge group, but chiral under the additional U(1) gauge symmetry).  Other $Z$-prime models have been considered in the context of asymptotic safety~\cite{Wang:2015sxe,Bause:2021prv,Chikkaballi:2022urc}, but also involving less than minimal sectors of additional matter and Higgs fields.   Our first motivation in the present work is to consider two $Z$-prime scenarios that are quite minimal in that the only fermions needed for the cancellation of anomalies are three generations of right-handed neutrinos.   The additional U(1) symmetries we consider are U(1)$_{B-L}$ and U(1)$_\chi$, where the latter is a well-known U(1) symmetry that commutes with SU(5) in SO(10) unified models~\cite{Carena:2004xs}.   These theories, which we will refer to as the B-L and $\chi$ models,  will be defined precisely in Sec.~\ref{sec:models}.

Our other motivation for studying simple $Z$-prime models is an interesting observation that the top-bottom mass difference might be understood as a consequence of their differing gauge charges in the standard model, which affect the values of their Yukawa couplings at nontrivial UV fixed points~\cite{Eichhorn:2017ylw,Eichhorn:2018whv}.   The larger fixed point value of the top quark Yukawa coupling is then the origin of the top-bottom mass splitting observed in the low-energy theory.   This is a conceptually appealing picture.   One expects that the differing gauge couplings between the right-handed top and bottom quarks due to new Abelian gauge symmetries will impact this conclusion and, conversely, may lead to restrictions on the new physics.  Even in the case of $B-L$ gauge symmetry, which does not distinguish top from bottom directly, there is a new effect due to the kinetic mixing of the $B-L$ gauge boson with hypercharge, which introduces effective couplings to the new gauge field that distinguish top from bottom.   We will constrain the two minimal models of interest by focusing on the most predictive possibility that both the top and bottom Yukawa couplings reach nontrivial UV fixed points.

Our paper is organized as follows.   In Sec.~\ref{sec:models}, we define the particle content and gauge charges of the two $Z$-prime models of interest.  We study the ultraviolet fixed points for the gauge and Yukawa couplings in Sec.~\ref{sec:fixedpts}, taking into account kinetic mixing between the Abelian gauge fields.   In Sec.~\ref{sec:conseq}, we discuss the consequences on viable model parameter space that follows from the requirement that desired nontrivial UV fixed points are reached.  In Sec.~\ref{sec:conc}, we summarize our conclusions.

\section{Models}\label{sec:models}
The models we consider are notable for their minimality. The gauge group is $G_{\rm SM} \times$~U(1)$_x$, where $G_{\rm SM}$ is the standard model gauge group and U(1)$_x$ is either U(1)$_{B-L}$ or U(1)$_\chi$. 
\begin{table}[h]
    \centering
    \begin{tabular}{cccc|cc}
        \hline\hline
        & \hspace{0.5em} SU(3)$_C$ \hspace{0.5em} & \hspace{0.5em} SU(2)$_W$ \hspace{0.5em} & \hspace{0.5em} U(1)$_Y$ \hspace{0.5em} &
        \hspace{0.5em} U(1)$_{B-L}$ \hspace{0.5em} & \hspace{0.5em} U(1)$_\chi$ \hspace{0.5em} \\
        \hline
        $q_L$  & 3 & 2 & 1/6 & 1/3 & 1/3  \\
        $u_R$   & 3 & 1 & 2/3 & 1/3 & -1/3 \\
        $d_R$ & 3 & 1 & -1/3 & 1/3 & 1 \\
        $\ell_L$ & 1 & 2 & -1/2 & -1 & -1 \\
        $e_R$ & 1 & 1 & -1 & -1 & -1/3 \\
        $\nu_R$ & 1 & 1 & 0 & -1 & -5/3 \\
        \hline\hline
    \end{tabular}
    \caption{Charge assignments for a single generation of fermions in the two models of interest.}
    \label{tab:one}
\end{table}
The subscripts on the U(1) factors are determined by the charge assignments of the fermions in the theory; we include only the three generations of standard model fermions with a right-handed neutrino $\nu_R$ included in each.  The charge assignments are shown in Table~\ref{tab:one}.

These assignments render each theory free of gauge and gravitational anomalies~\cite{Carena:2004xs}.   Notice that the U(1)$_\chi$ charges for the fields $\ell_L$ and $d_R^c$ are identical, where the superscript $c$ indicates charge conjugation; the same is true for $q_L$, $u_R^c$ and $e_R^c$.  This reflects the fact that U(1)$_\chi$ commutes with the familiar SU(5) gauge group from grand unified theories (GUTs), which implies that the components of any irreducible SU(5) representation  (in this case, the  ${\bf \overline{5}}$ and ${\bf 10}$) will all have the same U(1)$_\chi$ charges.  In addition,  if the standard model Higgs doublet $H$ is given the charges shown in Table~\ref{tab:two}, then the symmetries allow Yukawa couplings that lead to Dirac mass terms for all the fermions.   
\begin{table}[b]
    \centering
    \begin{tabular}{cccc|cc}
        \hline\hline
        & \hspace{0.5em} SU(3)$_C$ \hspace{0.5em} & \hspace{0.5em} SU(2)$_W$ \hspace{0.5em} & \hspace{0.5em} U(1)$_Y$ \hspace{0.5em} &
        \hspace{0.5em} U(1)$_{B-L}$ \hspace{0.5em} & \hspace{0.5em} U(1)$_\chi$ \hspace{0.5em} \\
        \hline
        $H$  & 1 & 2 & 1/2 & 0 & -2/3  \\
        $\phi$   & 1 & 1 & 0 & 1 & 1 \\
        \hline\hline
    \end{tabular}
    \caption{Charge assignments for the scalar sector in the two models of interest.}
    \label{tab:two}
\end{table}
While these theories can accommodate all standard model fermion masses without the need for any higher-dimension operators, they do not by themselves provide an origin for the hierarchy in fermion masses, including the smallness of neutrino masses relative to the charged fermions.   Those issues must be addressed by other physics.  While we only take into account the top-bottom mass splitting, asymptotic safety may play an important role in determining neutrino masses~\cite{Kowalska:2022ypk,deBrito:2025ges}, as well as the overall pattern of fermion masses and mixing angles~\cite{Alkofer:2020vtb,Eichhorn:2025sux}.  A study of the full flavor structure of the theories defined in Table~\ref{tab:one} is beyond the scope of the present work.

Finally, we note that the vacuum expectation value (vev) of the Higgs fields $H$ does not break the U(1)$_{B-L}$ symmetry, and only breaks the U(1)$_\chi$ symmetry at the electroweak scale $v \approx 246$~GeV, which is likely too low given typical phenomenological constraints~\cite{Carena:2004xs}.  Hence we assume that a standard model gauge singlet scalar $\phi$ is present in each theory with charge $+1$ under the additional U(1) symmetry of interest.  With a suitable potential, this allow the symmetry breaking scale of the new Abelian gauge symmetry to be raised and to remain in harmony with experimental bounds.

\section{Fixed Points}\label{sec:fixedpts}
The UV fixed points for the theories defined in the previous section are determined by the point in parameter space where the beta functions vanish.   At one-loop order, the beta functions for the gauge couplings are independent of the Yukawa couplings and couplings in the scalar potential, and may be considered separately.   The beta functions for the Abelian couplings may be written
\begin{equation}
\beta(g_i) = \frac{1}{(4 \pi)^2} \hat{\beta}^{(1)}(g_i) \,\,\, ,
\end{equation}
where
\begin{align}
\hat{\beta}^{(1)}(g_1) &= \frac{41}{10} \, g_1^3 - \theta(\mu-M_{\rm Pl}) \, \hat{f}_g \,g_1 \,\,\, ,  \label{eq:gbeta1} \\
\hat{\beta}^{(1)}(g) &= B_1 \, g^3 + \frac{41}{6} g \, \tilde{g}^2 + B_2 \, g^2 \tilde{g} - \theta(\mu-M_{\rm Pl}) \, \hat{f}_g \,g \,\,\, ,
\label{eq:gbeta2} \\
\hat{\beta}^{(1)}(\tilde{g}) &= B_3 \, g\, g_1^2 + B_4 \, g \, \tilde{g}^2 +\frac{41}{5} g_1^2 \, \tilde{g} +B_5 \, g^2 \tilde{g}+\frac{41}{6} \tilde{g}^3 - \theta(\mu-M_{\rm Pl}) \, \hat{f}_g \, \tilde{g} \label{eq:gbeta3}  \,\,\, .
\end{align}
where  $B^{B-L}_i = \{11, \, \frac{32}{3}, \, \frac{32}{5},\, \frac{32}{3}, \,11\}$ and $B_i^\chi = \{\frac{497}{27}, \, -\frac{4}{9}, \, -\frac{4}{15},\, -\frac{4}{9}, \,\frac{497}{27}\}$ are the coefficients for the  U(1)$_{B-L}$ model and for the U(1)$_\chi$ model, respectively.  Here, $g$ is the gauge coupling for the new U(1) gauge group and $g_1$ is the hypercharge gauge coupling in GUT normalization. Further, we have defined $\hat{f}_g \equiv (4 \pi)^2 f_g$ following the convention of Ref.~\cite{Boos:2022jvc}.  The parameter $\tilde{g}$ defines the kinetic mixing between the two Abelian gauge fields.   Writing the kinetic terms for the gauge fields in the matrix form
\begin{equation}
{\cal L} = -\frac{1}{4} (G^{-2})_{AB} F^A_{\mu\nu} F^{B\, \mu\nu} \,\,\, ,
\end{equation}
where the labels $A$ and $B$ run over the hypercharge and the new Abelian gauge group, then on may choose a field basis where $G$ has an upper-triangular form
\begin{equation}
G = \left(\begin{array}{cc} g_Y & \tilde{g} \\ 0 & g \end{array}\right) \,\,\, .
\end{equation}
The beta functions shown in Eqs.~(\ref{eq:gbeta1})--(\ref{eq:gbeta3}) were computed using the software PyR@TE~3~\cite{pyrate3}, which assumes this convention and encodes approach of Ref.~\cite{Poole:2019kcm}.  At lowest-order, a universal gravitational contribution to the running of gauge coupling matrix $G^2$ that is proportional to $f_g$ leads to the gravitational corrections in Eqs.~(\ref{eq:gbeta1})-(\ref{eq:gbeta3}) that are linear in $g_1$, $g$ and $\tilde{g}$, respectively.  Similar assumptions have been made in other treatments of kinetic mixing in the phenomenological literature on asymptotic safety (for example in Refs.~\cite{Wang:2015sxe,Chikkaballi:2022urc}), and these are consistent with the linear corrections obtained in functional renormalization group studies~\cite{Folkerts:2011jz,Eichhorn:2016esv,Christiansen:2017gtg,Eichhorn:2017eht}.

It is natural to divide the set of possible fixed point into two cases, namely where $g_{1*}$ is non-zero or zero, respectively.   With the choice,
\begin{equation}
\hat{f}_g = \frac{41}{10} g_1(M_{\rm Pl})^2 \,\,\, ,
\label{eq:fgcritv}
\end{equation}
then $g_{1*}=g_1(M_{\rm Pl})$;  for larger choices of $\hat{f}_g$, $g_1$ will flow to zero in the UV, {\it i.e.}, $g_{1*}=0$.   The fixed points for $g$ and 
$\tilde{g}$ differ in each of these cases:

{\it Nontrivial $g_1$ fixed point}.   In the case where $g_{1*}$ is nonvanishing, Eqs.~(\ref{eq:gbeta1})--(\ref{eq:gbeta3}) imply UV fixed points at the 
origin, $(g_*,\tilde{g}_*) = (0,0)$, and at
\begin{align}
g_* & = (B_1 - \frac{25}{246} B_3^2)^{-1/2}\, \hat{f_g}^{1/2} \\
\tilde{g}_* &= -\frac{5}{41} B_3 \, g_* \label{eq:theline}
\end{align}
where we have used the fact that $B_1=B_5$, $B_2=B_4$ and $B_3 = (3/5)\, B_2$ in both models.   These reduce to the points 
\begin{align}
&(g_*,\tilde{g}_*)_{B-L} = \left(\frac{\sqrt{123}}{29} \hat{f}_g^{1/2},\,\, -\frac{32}{29}\sqrt{\frac{3}{41}}\hat{f}_g^{1/2}\right) \,\, ,
\label{eq:pointBmL} \\
&(g_*,\tilde{g}_*)_{\chi} = \left(3 \sqrt{\frac{123}{20369}} \hat{f}_g^{1/2},\,\, 4\sqrt{\frac{3}{835129}}\hat{f}_g^{1/2}\right) \,\, ,
\label{eq:pointChi} 
\end{align}
where the subscript on the left-hand-side indicates the model in question.
\begin{figure}[!htb]
\centering
\includegraphics[width=0.5\textwidth]{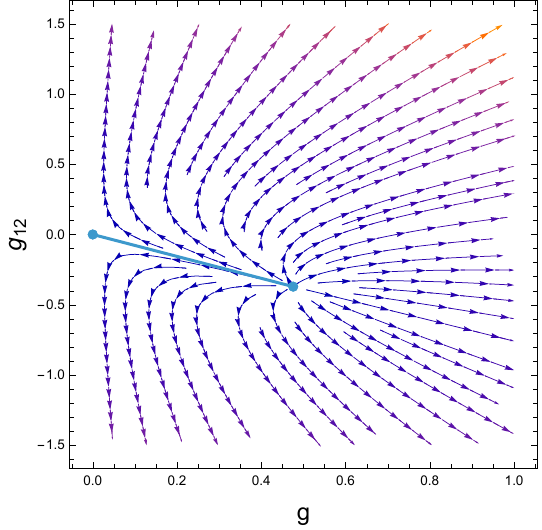}
\caption{Fixed points (solid dots) in the $(g,\tilde{g})$ plane for $g_{1*} = g_1(M_{\rm Pl})$ in the B-L model. The solid line connecting the fixed points is the UV critical surface defined by Eq.~(\ref{eq:theline}).}
\label{fig:BmLnontrivial}
\end{figure}
For the purpose of illustration we show the renormalization group flow in the vicinity of the fixed points for the $B-L$ model in 
Fig.~\ref{fig:BmLnontrivial}.   The fixed point away from the origin is entirely repulsive, which makes it the most interesting from the perspective of predictivity; low-energy coupling values must be chosen to reach that point exactly or the couplings will flow either to a Landau pole, or towards the fixed point at the origin.   The latter will happen for coupling values that lie on the line defined by Eq.~(\ref{eq:theline})  that connects the two fixed points.  The qualitative features of the flows in the $\chi$ model are similar, aside from the slope of the line connecting the two fixed points, so we do not display the corresponding figure.

{\it Trivial $g_1$ fixed point}.  The hypercharge gauge coupling flows to a Gaussian fixed point for values of $\hat{f}_g > \hat{f}_g^{crit}$, where
\begin{equation}
\hat{f}_g^{crit}  \approx 1.5473 \,\,\, .
\label{eq:fcrit}
\end{equation}
This numerical value follows from Eq.~(\ref{eq:fgcritv}) and $\alpha_1^{-1}(M_{Pl})$ as determined from the data given in Appendix~\ref{sec:appendix}. In this case, the fixed points in the $g$-$\tilde{g}$ plane fall in an ellipse defined by the equation
\begin{equation}
\frac{B_1}{\hat{f}_g}\, g^2 + \frac{B_2}{\hat{f}_g}\, g \tilde{g} + \frac{41/6}{\hat{f}_g}\, \tilde{g}^2 =1  \,\,\,.
\label{eq:theellipse}
\end{equation}
The precise orientation of the ellipse depends on the model; the angle $\theta$  between the semi-minor and the $x$-axis is given by
\begin{equation}
\tan 2\theta = \frac{B_2}{B_1 - 41/6} \,\,\, .
\end{equation}
This yields $\theta \approx 34.3^\circ$ ($\theta \approx -1.1^\circ$) in the B-L ($\chi$) model, independent of $\hat{f}_g$.
\begin{figure}[!htb]
\centering
\includegraphics[width=0.5\textwidth]{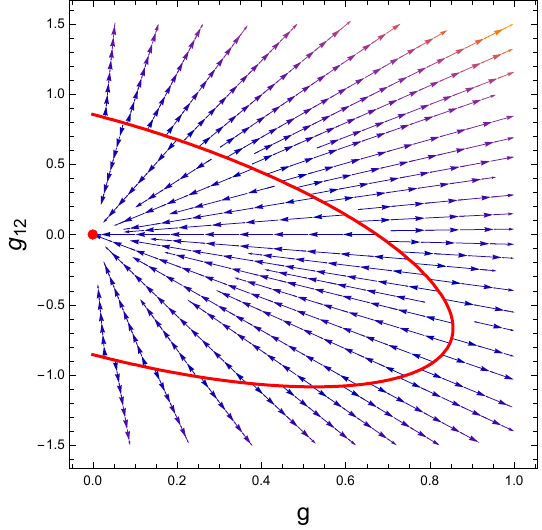}
\caption{Fixed points (solid dot and ellipse) in the $(g,\tilde{g})$ plane for $g_{1*} = 0$ in the B-L model.  For the purpose of illustration, we 
have chosen $\hat{f}_g=5$.}
\label{fig:BmLtrivial}
\end{figure}
Coupling values on the interior of the ellipse will eventually flow to the trivial fixed point at the origin.   We do not display the figure for the $\chi$ model, which has the same qualitative features as the B-L model aside from the rotation of the ellipse of fixed points.

In the present work, we focus on the most predictive possibilities, namely the choices where $g$ and $\tilde{g}$ flow to the nontrivial fixed point in Fig.~\ref{fig:BmLnontrivial}, or to the any on the ellipse of fixed points shown in Fig.~\ref{fig:BmLtrivial}, as well as the analogous choices in the $\chi$ model.  In addition, we look at the possibility that the top-bottom quark mass splitting might be understood in terms of the different fixed point values of their Yukawa couplings.  The beta functions are given by
\begin{align}
\beta^{(1)}(y_t) &=\frac{9}{2}\, y_t^3+\frac{3}{2}\, y_t \, y_b^2 -\frac{17}{20} \,g_1^2\, y_t -\frac{2}{3} \, g^2 \, y_t  +C_1 \, g \, \tilde{g} \,y_t 
-\frac{17}{12} \,\tilde{g}^2\, y_t-\frac{9}{4} \, g_2^2\, y_t \nonumber \\ & -8 \, g_3^2\, y_t - f_y\, y_t  \, ,  \label{eq:yt} \\
\beta^{(1)}(y_b) &=\frac{9}{2}\, y_b^3+\frac{3}{2}\, y_b \, y_t^2 -\frac{1}{4} \,g_1^2\, y_b +C_2\, g^2 \, y_b +C_3 \, g \, \tilde{g} \, y_b
-\frac{5}{12} \,\tilde{g}^2\, y_b-\frac{9}{4} \, g_2^2\, y_b \nonumber \\
&-8 \, g_3^2\, y_b - f_y \, y_b   \, , \label{eq:yb}
\end{align}
where $C^{B-L}_i = \{ -5/3,\,-2/3,\, 1/3\}$. and $C_i^\chi = \{1,\, -10/3,\, 5/3\}$ are the coefficients for the  $B-L$ and $\chi$ models, respectively, and we assume for simplicity that $y_b$ and $y_t$ are real.  Requiring that nontrivial fixed points are reached for both $y_b$ and $y_t$ leads to the following restriction on the fixed point values of the Yukawa couplings:
\begin{equation}
y_{t*}^2 - y_{b*}^2 = \frac{1}{5} \, g_{1*}^2 + \frac{1}{3}\, (C_2+2/3) \, g_*^2 - \frac{1}{3} (C_1-C_3) \, g_* \,\tilde{g}_* + \frac{1}{3} \, \tilde{g}^2_*. \,\,\, .
\label{eq:starsplit}
\end{equation}
Although we have omitted the $\tau$ Yukawa coupling in Eqs.~(\ref{eq:yt}) and (\ref{eq:yb}), we note that Eq.~(\ref{eq:starsplit}) is unchanged if those terms are included.  A particular choice of UV fixed points in the gauge sector will lead to a specific splitting between the fixed point values of the top and bottom Yukawa couplings; as we will see in the next section, these can then be use to determine consistency with the observed top and bottom quark mass splittings in the low-energy theory.

\section{Consequences of Asymptotic Safety}\label{sec:conseq}
\begin{figure}[t]
\centering
\includegraphics[width=0.7\textwidth]{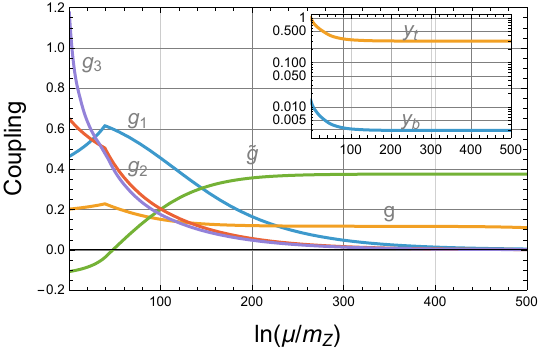}
\caption{Running of the couplings in the B-L model, in the case where $g_{1*} =0$, for the case where both $g$ and $\tilde{g}$ reach nontrivial fixed point values. The inset shows the top and bottom Yukawa couplings on a log scale so that the approach of $y_b$ to a nontrivial fixed point is easer to see.}
\label{fig:nicerun}
\end{figure}
We now discuss the consequences of asymptotic safety for the two models of interest, using the following algorithm: 

We choose an input value of the new U(1) gauge coupling $g$ at the renormalization scale $\mu=M_{\rm Pl}$, with $\tilde{g}$ then determined at the same scale by the UV critical surface constraint.  In the case where $g_{1*}$ is nonvanishing,  this is the line in the $g$-$\tilde{g}$ plane that connects the origin to the fixed point defined in Eqs.~(\ref{eq:pointBmL}) and (\ref{eq:pointChi}).   The case where $g_{1*}$ is nonvanishing is the most restrictive possibility since the value of $\hat{f}_g$ is fixed at the critical value given by Eq.~(\ref{eq:fcrit}), {\it i.e.}, $\hat{f}_g^{crit}  \approx 1.5473$.   Hence the nontrivial critical point that defines the end of this line has the numerical value $(g_*,\tilde{g}_*)_{B-L} = ( 0.4757, -0.3713)$ and $(g_*,\tilde{g}_*)_{\chi}=(0.2900 ,0.0094)$ in our two models, respectively; note that the line that defines the UV critical surface in the B-L case is displayed in Fig.~\ref{fig:BmLnontrivial}.  In the case where $g_{1*}=0$, there is more freedom since  $\hat{f}_g > \hat{f}_g^{crit}$, with the critical surface forming an ellipse in the $g$-$\tilde{g}$ plane defined by Eq.~(\ref{eq:theellipse}), including it's interior; the size of the ellipse is determined by the value of $\hat{f}_g$.   This critical surface is visible in the example displayed in Fig.~\ref{fig:BmLtrivial}.  

With these couplings fixed, the complete set of gauge couplings (including $g_2$ and $g_3$) can be run to determine their fixed point values $g_{i*}$.  Of course, we can only approximate the $\mu \rightarrow \infty$ limit by evaluating couplings at a very high scale $\mu_{\rm max}$.  We typically define $\mu_{\rm max}$ such that
\begin{equation}
\ln\left(\frac{\mu_{\rm max}}{M_{\rm Pl}}\right) = 500 \,\,\, .
\end{equation}
This choice is typically high enough to capture the approach of the theory to its fixed point values; this is illustrated in Fig.~\ref{fig:nicerun}.  At the scale $\mu_{\rm max}$, with the gauge couplings now specified, the fixed point values of the top and bottom Yukawa couplings are determined by the value of $\hat{f}_y$.  We choose that parameter so that the correct bottom quark Yukawa coupling is obtained at the electroweak scale.   The value of the top quark Yukawa coupling at the electroweak scale is now fixed and can be compared to its experimental value.   We provide the numerical input values of the gauge and Yukawa couplings used in this analysis in Appendix~\ref{sec:appendix}.  By scanning over the allowed range of $g(M_{\rm Pl})$, we can determine whether an acceptable value of $y_t$ is obtained.  
\begin{figure}[t]
\centering
\includegraphics[width=0.49\textwidth]{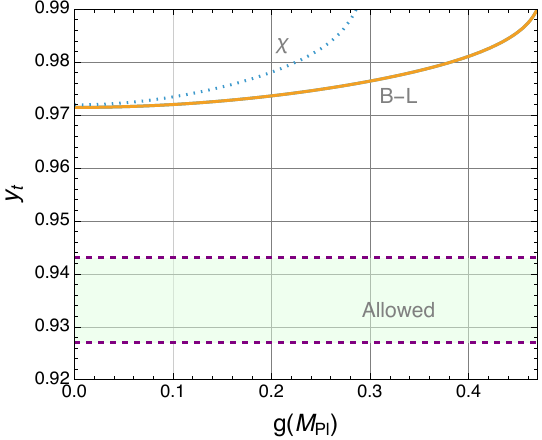}
\includegraphics[width=0.49\textwidth]{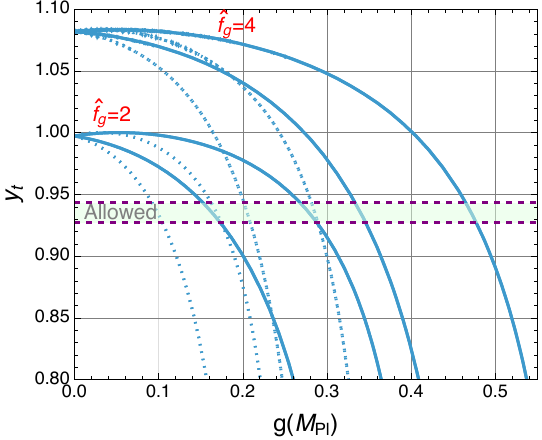}
\caption{Predictions for the top quark Yukawa coupling renormalized at the scale $m_{t}$.  The solid (dotted) lines refer to the $B-L$ ($\chi$) model.  The shaded band between the dashed lines is the 2 standard deviation experimentally allowed range.}
\label{fig:ytplots}
\end{figure}

\begin{figure}[]
\centering
\includegraphics[width=0.45\textwidth]{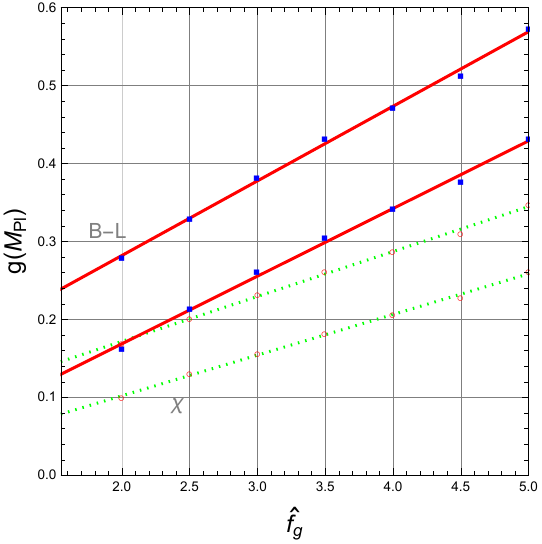}
\caption{Value of $g(M_{\rm Pl})$ leading to the correct top and bottom quark masses as a function of the parameter $\hat{f}_g$, in the two
models of interest in the case where $g_{1*}=0$. The lines shown are linear fits which each have a goodness-of-fit parameter $R$ that exceeds 0.998.}
\label{fig:rightg}
\end{figure}

The output results of this algorithm are shown in Fig.~\ref{fig:ytplots}.    In the subfigure on the left,  we scan over the full allowed range of $g(M_{\rm Pl})$; the endpoint of each curve at the top of the plot coincides with one end of the critical surface line, with the solid (dotted) lines corresponding to the B-L ($\chi$) model.  In both case, the predicted value of $y_t$ is between $3$\% and $8$\% larger than the $2\sigma$ upper limit on the Yukawa coupling renormalized at the scale $m_t$, indicated by the horizontal shaded band.   Hence, this solution is excluded, given our assumption about the form of the gravitational corrections to the renormalization group equations.  On the other hand, the subfigure on the right displays the results from scanning over the ellipse of fixed points.  This is a subset of the full UV critical surface, which includes the interior of the ellipse, but one which is the most restrictive.   In this case, the solid (dotted) lines again correspond to the predictions for $y_t$ in the B-L ($\chi$) models.   The lines come in pairs since the ellipse provides two solutions for $\tilde{g}(M_{\rm Pl})$ for each choice of $g(M_{\rm Pl})$.  Each of these solutions for $g(M_{\rm Pl})$ grow monotonically as a function of $\hat{f}_g$ over the range we studied numerically.   The values of $g(M_{\rm Pl})$ that are consistent with the desired fixed point solutions for the top and bottom quark Yukawa couplings as a function of $\hat{f}_g$ are shown in Fig.~\ref{fig:rightg}.  The lines shown are linear fits to the numerical data points, which provide an accurate description of the dependence on $\hat{f}_g$;  each fit shown has a goodness-of-fit parameter $R$ greater than $0.998$.   We note that a functional renormalization group analysis that predicts $\hat{f}_g$, assuming the particle content of these models, would lead to predictions for both $g$ and $\tilde{g}$ that can then be run to observable energies; as we have noted earlier, however, this approach involves significant uncertainties.
Finally, we note that we have checked that the results shown are not affected by inclusion of the tau lepton Yukawa coupling, which runs to a Gaussian fixed point given the values of $f_y$ that are relevant in the previous analysis.   Our results are also not affected by couplings in the scalar sector.  The running in that sector is controlled by a different gravitational parameter, $f_\lambda$, that can be chosen so that coupling flow to Gaussian fixed points, similar to the gauged baryon number model of Refs.~\cite{Boos:2022jvc,Boos:2022pyq}.

\section{Conclusions}\label{sec:conc}

Although it was proposed originally in the context of gravity alone~\cite{weinberg0}, asymptotic safety provides a useful paradigm for constraining possible extensions of the standard model~\cite{Bond:2017wut,Wang:2015sxe,Grabowski:2018fjj,Hiller:2019mou,Kowalska:2020zve,Kowalska:2020gie,Bause:2021prv,Chikkaballi:2022urc,Boos:2022jvc,Boos:2022pyq,Chikkaballi:2025pnw}.   In the present work, we considered two, minimal Z-prime models -- minimal in the sense that no additional matter was required to cancel gauge anomalies other than a right-handed neutrino for each standard model fermion generation.   Adopting a now standard procedure for approximating the gravitational corrections to the renormalization group running of couplings above the Planck scale, we considered the implications of the requirement that some of the gauge couplings, as well as the top and bottom quark Yukawa couplings, reach nontrivial ultraviolet (UV) fixed points.   The results we obtained were in some ways qualitatively similar to those found in a different, less minimal Z-prime model, the model of gauged baryon number considered in Ref.~\cite{Boos:2022jvc,Boos:2022pyq}.   Two classes of fixed points were considered, depending on whether the hypercharge coupling $g_1$ reaches a Gaussian or a nontrivial fixed point in the UV.  In either case, the constraints on the UV behavior of the new gauge coupling and its kinetic mixing parameter relates them, so that the kinetic mixing parameter is not a free parameter in the low-energy theory.   This is a small but phenomenologically meaningful reduction in the model parameter space.  Adding the constraint that the top and bottom quark Yukawa couplings reach nontrivial UV fixed points selects the trivial $g_1$ fixed point solutions as preferred.  We then identified viable points in model parameter space where the gauge coupling and kinetic mixing are effectively determined by the value of the gravitational correction parameter for the gauge couplings, $f_g$, something which might separately be obtained through a nonperturbative calculation. 

The limitation of the present work is that we have not eliminated all the model building freedom.  We have not exhausted the combinatorial possibilities for the sets of couplings that reach either trivial or non-trivial fixed points, nor have we fixed the gravitational parameters from a first principles calculation.    It may be worthwhile for future work to consider the latter issue using functional renormalization group methods while assuming the specific field content that defines these minimal models.  If a reduction in the uncertainty of the gravitational parameters can be achieved, one would then determine the new gauge sector couplings that could be probed directly if the scale of the U(1) symmetry breaking falls within the reach of future collider experiments.

\begin{acknowledgments} 
CDC thanks the NSF for support under Grant No. PHY-2112460 and No. PHY-2411549.
\end{acknowledgments}

\appendix
\section{Input parameters} \label{sec:appendix}
The numerical results shown in Sec.~\ref{sec:conseq} assume the following values of standard model masses and couplings, based on data compiled by the Review of Particle Physics~\cite{ParticleDataGroup:2024cfk}.  None of our results depend strongly on the precise choices of these numerical inputs (for example, a global fit value versus a specific experimental measurement):

$\bullet$ {\it Top Yukawa coupling}.  We assume the top quark pole mass $m_t = 172.56 \pm 0.31$~GeV, yielding 
$m_t^{\overline{MS}} = 162.76 \pm 0.65$~GeV.  This gives the value assumed in our numerical analysis
\begin{equation}
y_t^{\overline{MS}}(m_t) = 0.935 \pm 0.004 \,\, ,
\end{equation}
following from the electroweak scale $v=(\sqrt{2} \, G_F)^{-1/2} = 246.22$~GeV.

$\bullet$ {\it Bottom Yukawa coupling}.  We use $m_b^{\overline{MS}} (m_b) = 4.183 \pm 0.007$~GeV, corresponding to 
$m_b^{\overline{MS}} (m_t) = 2.36 \pm 0.06$~GeV.   With the value of $v$ quoted above, we obtain
\begin{equation}
y_b^{\overline{MS}}(m_t) = 0.0136 \pm 0.0003 \,\, .
\end{equation}

$\bullet$  {\it Standard model gauge couplings}.  We assume
\begin{align}
\alpha_1^{-1}(m_Z) =& 59.005\pm 0.005 \nonumber \\
\alpha_2^{-1}(m_Z) =& 29.589\pm 0.005 \nonumber \\
\alpha_3^{-1}(m_Z) =& 8.475 \pm 0.065 
\end{align}
These are run to $m_t$, to obtain the boundary conditions for the renormalization group analysis described in the text.

\end{document}